\documentclass[a4paper]{article}

\usepackage{INTERSPEECH2022}
\usepackage{graphicx}
\usepackage{amsmath}
\usepackage{bbm}
\usepackage{wasysym}
\usepackage{multirow}
\usepackage{algpseudocode}
\usepackage{float}
\usepackage{url}
\usepackage{xcolor}

\newcommand{\X}{\ensuremath{\mathcal{X}}}
\newcommand{\Z}{\ensuremath{\mathcal{Z}}}
\newcommand{\Y}{\ensuremath{\mathcal{Y}}}
\newcommand{\x}{\ensuremath{\mathbf{x}}}
\newcommand{\seqit}[1]{\bar{#1}}
\newcommand{\sy}{\seqit{y}}
\newcommand{\sz}{\seqit{z}}
\newcommand{\sx}{\seqit{\x}}
\newfont{\msym}{msbm10}
\newcommand{\reals}{\mbox{\msym R}}

\newcommand{\lstm}{DDKtor-LSTM}
\newcommand{\cnn}{DDKtor-CNN}

\newcommand{\minitab}[2][l]{\begin{tabular}{#1}#2\end{tabular}}

\title{DDKtor: Automatic Diadochokinetic Speech Analysis}
\name{Yael Segal$^1$, Kasia Hitczenko$^2$, Matthew Goldrick$^3$, Adam Buchwald$^4$, Angela Roberts$^5$, and Joseph Keshet$^1$}
\address{
  $^1$Faculty of Electrical and Computer Engineering, Technion--Israel Institute of Technology, Israel\\
  $^2$Laboratoire de Sciences Cognitives et Psycholinguistique, D\'epartement d’Etudes Cognitives, ENS, EHESS, CNRS, PSL University, France\\
  $^3$Department of Linguistics, Northwestern University, IL, USA\\
  $^4$Department of Communicative Sciences and Disorders, New York University, NY, USA\\
  $^5$Department of Computer Science and School of Communication Sciences and Disorders, University of Western Ontario, Ontario, Canada}
\email{segal.yael@campus.technion.ac.il, kasia.hitczenko@ens.psl.eu, matt-goldrick@northwestern.edu, buchwald@nyu.edu, angela.roberts@uwo.ca, jkeshet@technion.ac.il}

\begin{document}

\maketitle

\begin{abstract}
  Diadochokinetic speech tasks (DDK), in which participants repeatedly produce syllables, are commonly used as part of the assessment of speech motor impairments. These studies rely on manual analyses that are time-intensive, subjective, and provide only a coarse-grained picture of speech. This paper presents two deep neural network models that automatically segment consonants and vowels from unannotated, untranscribed speech. Both models work on the raw waveform and use convolutional layers for feature extraction. The first model is based on an LSTM classifier followed by fully connected layers, while the second model adds more convolutional layers followed by fully connected layers. These segmentations predicted by the models are used to obtain measures of speech rate and sound duration. Results on a young healthy individuals dataset show that our LSTM model outperforms the current state-of-the-art systems and performs comparably to trained human annotators. Moreover, the LSTM model also presents comparable results to trained human annotators when evaluated on unseen older individuals with Parkinson's Disease dataset.
\end{abstract}


\section{Introduction}
\label{sec:intro}
Diadochokinetic (DDK) speech tasks are commonly used by clinicians and researchers as part of the assessment of speech motor impairments \cite{kent1987maximum,nishio2006comparison}. In the alternating motion rate (AMR) version of this task, participants repeatedly produce particular syllables as quickly and accurately as possible (e.g., pa-pa-pa..., ta-ta-ta..., or ka-ka-ka...). In the sequential motion rate (SMR) task, a syllable sequence is repeatedly produced (e.g., pa-ta-ka-pa-ta-ka...). These tasks help clinicians evaluate the patient's speech motor control and ability to make rapidly alternating speech movements. These tasks have been shown to be useful for impairment detection, differential diagnosis, and course monitoring and have, consequently, become a part of many speech/neurological assessments \cite{baken2000speech,gadesmann2008}.

For a measurement that has been proven to be important in patient care, the outcome measures in use are surprisingly basic, with clinicians evaluating the patient impressionistically and/or counting how many syllables the patients were able to produce in a certain amount of time and comparing that against established norms. While these measures are easy for clinicians to obtain, impressionistic evaluation is inherently subjective, and previous work has suggested that both of these measures have relatively low inter- and intra-rater reliability \cite{gadesmann2008} (but see \cite{Karlsson2020}). These measures also provide a highly impoverished picture of speech. The acoustics of DDK productions provide information about the temporal and spectral properties of syllables and individual speech sounds as well as more complex measures of speech rate (e.g., variance, irregularity). While such measures have been studied in research settings by manually annotating the speech \cite{ziegler2002task}, this approach is clearly impractical in a clinical setting. 

Automated, objective measurement of more detailed properties of untranscribed, unannotated DDK speech could provide a means of addressing these issues without placing additional burdens on valuable clinician time. Previous studies focused on automatic annotation of a single property. Signal processing methods \cite{rasanen2018pre} and deep learning methods utilizing convolutional neural networks (CNNs) \cite{wang2019deepddk, rozenstoks2019automated} have been used to segment syllables, allowing for automatic calculation of speech rate. Another line of works has been focused on measuring voice onset time (VOT). Monta\~na et al.  \cite{montana2018diadochokinesis} proposed an expert system that is based on temporal and spectral features, and Arias-Vergara et al. \cite{gru} proposed to use a deep learning approach that utilized a bi-directional recurrent neural network on manually extracted time and spectral acoustic features. As far as we know, only Novotn{\`y} et al. \cite{novotny2014automatic} suggested an automated segmentation of both VOT and vowels, by using a representation similar to \cite{montana2018diadochokinesis} without deep learning.

In this paper, we suggest two novel deep learning models for automatic segmentation of unannotated diadochokinetic speech. Our models are unique as they allow an accurate segmentation of both vowels and VOTs, with the flexibility of using variable-length processing window and by working directly on the raw waveform. These architectures eliminate the need of a restrictive representation and translate to a more accurate measurement of the speaking rate. 
Using DDK samples from both healthy individuals and individuals with Parkinson's Disease (PD), we show that our model outperforms the current state-of-the-art and performs comparably to trained human annotators. The implementation of our models is available at:
\url{https://github.com/MLSpeech/DDKtor}.


\section{Model}
\label{sec:model}

In the DDK task, we are provided with an \emph{audio signal} containing an alternating sequence of a positive-lag VOT (for each voiceless stop consonant) followed by the vowel  $\langle a \rangle$. Our goal is to segment the audio signal according to three acoustic objects: VOT, vowel, and other/silence. Hence, the input to our models is the raw audio and the output is a sequence of objects and their timings. 

\begin{figure*}[h]
 \centering
\includegraphics[width=1\linewidth, height=0.3\textwidth]{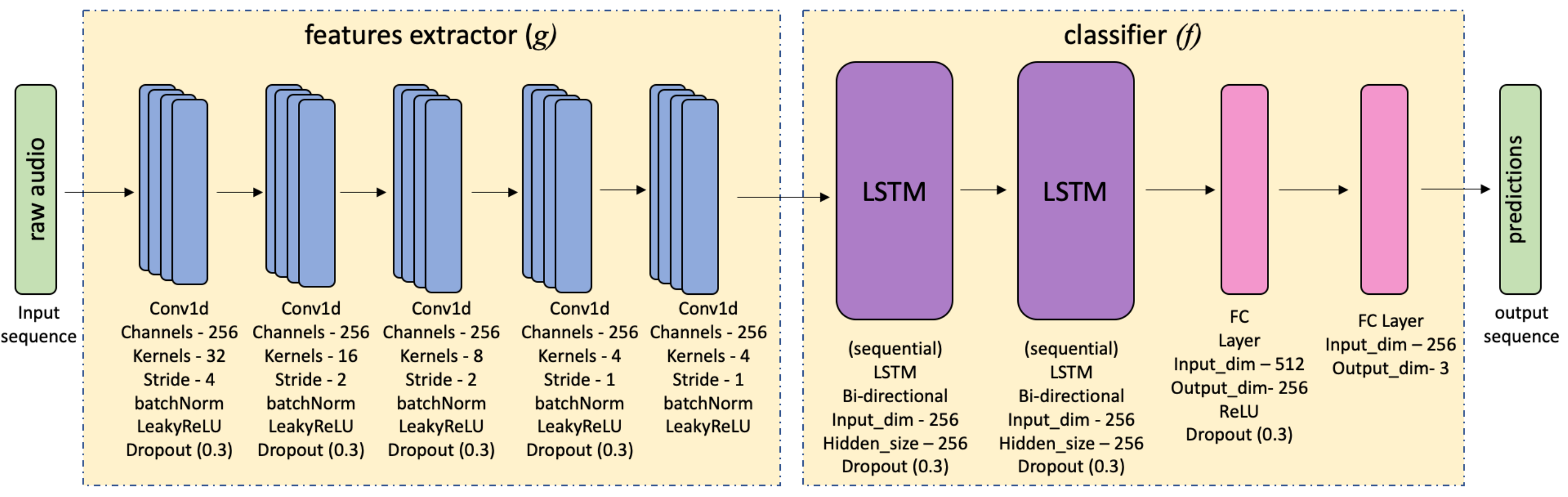}
\includegraphics[width=1\linewidth, height=0.3\textwidth]{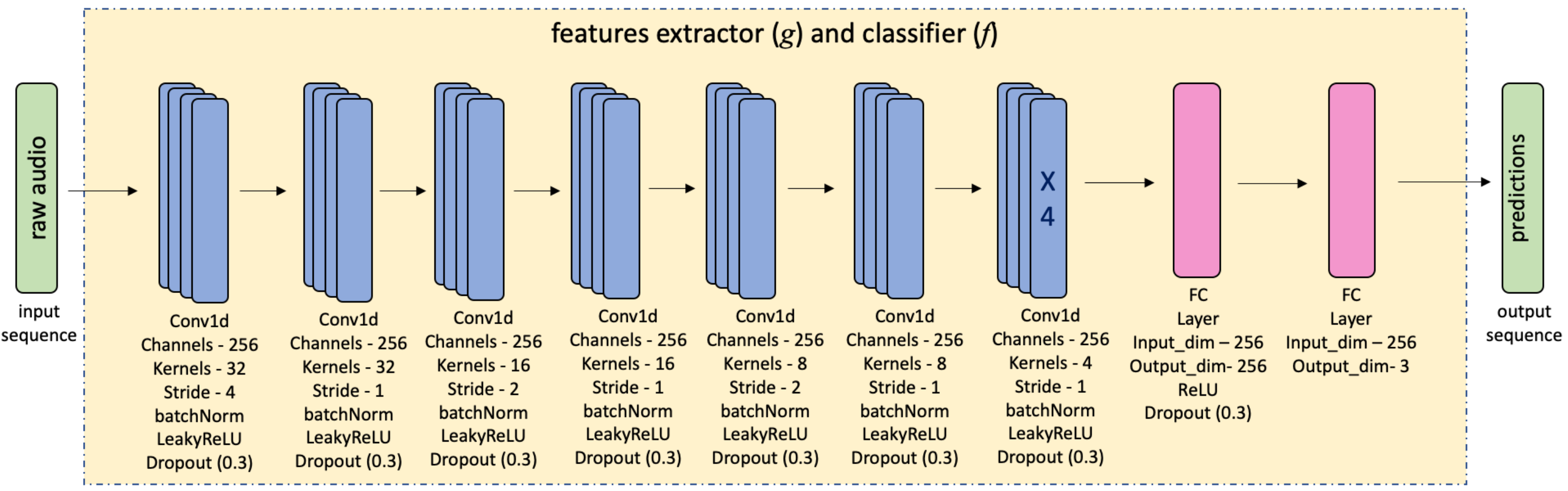}

\vspace{-.4cm}
 \caption{Features extractor and classifier architecture.  \lstm~(top): 5 layers of CNN as features extractor, 2 layers LSTM followed by 2 fully-connected (FC) layers as classifier. \cnn~(bottom): 10 layers of CNN followed by 2 FC layers as feature extractor and classifier.}
 \label{fig:architectures}
\end{figure*}

For raw audio of duration T, we denote the input sequence of samples by
$\sx  = (x_1,\ldots,x_T)$, where $x_t\in\X$ for all $1 \leq t \leq T$ and $\X \subset \reals $. We represent the output as a sequence of 1 milliseconds frames, $\sy = (y_1,\ldots,y_K)$, where there are $K$ frames, and $y_k\in \Y=\{\text{VOT, vowel, other}\}$ for  $1 \leq k \leq K$. Note that the signal duration $T$ and the number of output frames $k$ can vary from one signal to another, and thus these quantities are not fixed. 

Our models are composed of two functions. The \emph{features extraction function} $g: \X\to \Z^K$ is a function from the domain of $\X$ to the domain of an abstract latent representation $\Z \subseteq \reals^N$. Namely, the features extraction function generates a sequence of embedding vectors $\sz  = (z_1,\ldots,z_K)$. Each vector $z_k\in\Z$ represents the acoustic content of the $k$-th frame. The feature vector sequence is then processed by a classification function that outputs a sequence of $K$ predictions. The \emph{classifier function} $f: \Z^K\to \Y^K$ from the domain of features vectors to the domain of target objects.

The functions $g$ and $f$ are implemented using deep learning functions by two models. The first model is depicted in the top panel of Figure~\ref{fig:architectures}. In this model, a convolutional neural network (CNN) is used as a features extraction function, $g$. It has five 1D convolutional layers with batch normalization, a leaky-ReLU activation function, and dropout between each layer. The features extractor output is then forwarded to the classifier $f$, composed of a two-layer bi-directional long short-term memory (LSTM) and two fully-connected (FC) layers. This model will be denoted as \emph{\lstm}. 

The second model is depicted in the bottom panel of Figure~\ref{fig:architectures}. In recent work, it has been suggested that CNN architectures are an effective replacement of recurrent models \cite{collobert2016wav2letter}, and we exploit this idea in our models. In our second model, a CNN architecture is used for the feature extraction function $g$ and the classifier function $f$. It has ten 1D convolutional layers with batch normalization, a leaky-ReLU activation function, and dropout between each layer. The output of the CNN is forwarded to two FC layers. This model will be denoted as \emph{\cnn}.

The parameters of both models are trained to minimize the cross-entropy loss function. For both models, a \emph{post-processing procedure} is used to convert the frame-based to segment-based predictions. First, we group together frames with the same object type.

Second, we mark short VOTs (less than 5 milliseconds), and short vowels (less than 20 milliseconds) as silence. .
Finally, we convert a short silence (less than 20 milliseconds) between two VOT segments to a single VOT segment.

\section{Datasets}
\label{sec:datasets}

In our experiments, we used two datasets of DDK productions in English. Both datasets were annotated for VOTs and vowel durations by two independent annotators. 
The first is called the \emph{Younger NT Adults} dataset, and it includes speech from the AMR and SMR subtasks for 92 neurotypical adult participants (mean age in their early twenties), collected in a laboratory environment as pre-test data in speech motor learning experiments \cite{buchwald2019, cheng2021}. The speech signals were sampled at 44.1 kHz with 16-bit resolution. Participants were randomly split into training (\emph{N} = 55, AMR $\sim$9 minutes, SMR $\sim$3 minutes), validation (\emph{N} = 18,  AMR $\sim$ 3 minutes, SMR $\sim$1 minutes), and test (\emph{N} = 19,  AMR $\sim$4 minutes, SMR $\sim$2 minutes).

To test the ability of the algorithm to generalize to laboratory speech from individuals with motor speech impairments, we used a second dataset called the \emph{Older PD Adults}.  This dataset contains the AMR and SMR subtasks from \emph{N} = 5 older adults with Parkinson's Disease (PD), aged 59-77 years old. These were selected from the Ontario Neurodegenerative Disease Research Initiative (ONDRI), a longitudinal, multi-site, observational cohort study, using a transdisciplinary approach to characterizing deep endophenotypes in neurodegenerative disorders and their relationship to cerebrovascular disease \cite{farhan2017,mclaughlin2021}. The speech signals were sampled at 44.1 kHz with 16-bit resolution. Manual analysis conducted prior to this study, focusing solely on the five selected individuals, had extracted for each speaker $\sim$5 seconds from the AMR subtask production for each VOT and $\sim$5 seconds from the SMR subtask (syllables were left intact). Note that this dataset was used only to evaluated our model and not for training.

\section{Experiments}
\label{sec:experiments}


\subsection{Details}

Both \lstm~and \cnn~were trained with a batch size of 32. We optimized parameters with the \emph{Adam} optimizer \cite{kingma2014adam} and a learning rate of 0.0001. The audio files were resampled at 16 kHz. Long audio files were divided into one-second segments. We utilized data augmentations to increase the robustness of the algorithm. We used the package \emph{WavAugment} \cite{wavaugment2020} to augment the data using: (i) clean speech; (ii) noisy speech\footnote{We added car-noise from \cite{varga1992noisex} package, which we found most similar to the air condition noise often evident in the datasets, although other noises work similarly.} with signal-to-noise ratio of 5, 10, 15 dB; and (iii) band-reject filtered speech (removing randomly selected spectral components). We also randomly shifted the starting frame of each one-second input for generating different inputs lengths. 

We compared our models against the state-of-the-art model Arias-Vergara et al. \cite{gru}. We trained it on \emph{Younger NT Adults} dataset, without using data augmentation, as it dramatically reduced performance. We also compared our model against  R\"{a}s\"{a}nen et al. \cite{rasanen2018pre}, which presents an algorithm for segmentation of syllables-like objects. Hence we compare it only for the Diadochokinetic speech rate task, using the best parameters they reported in their study ($f0$ = 8Hz, Q = 0.8, and $\delta$ = 0.01). Implementations of \cite{wang2019deepddk, rozenstoks2019automated, montana2018diadochokinesis, novotny2014automatic} were not available.

\subsection{Evaluation and Results}
 
We evaluate model performance against the gold standard: measures derived from manual annotations (we compare against one annotator; see the GitHub repository for results against the second annotator). As a benchmark of the performance, we also provide measures of inter-annotator agreement denoted as \emph{Annotators}. Below, we report results on the two unseen test sets we consider.

\subsubsection{Diadochokinetic speech rate}

DDK rate is defined as the number of syllables produced divided by total articulation time. That is, the time elapsed between the VOT onset of the first produced syllable and the vowel offset of the final produced syllable. Four different rates were calculated for each participant (one for each syllable of the AMR task; one for the SMR task).

We analyzed the decisions of \lstm~and \cnn~and found that both models sometimes merge two adjacent syllables (treated flap t's as a part of the vowel). To automatically account for this issue, we incremented the syllable count every time the duration of a predicted vowel was more than twice the participant's average vowel duration. 
The model of \cite{gru} sometimes skips VOTs. We made a similar adjustment by incrementing syllable count whenever the time elapsed between two VOTs was more than twice the participant's average inter-VOT duration. 

Additionally, note that because \cite{gru} estimates only VOT, it cannot calculate total articulation time (which requires a value for total syllable length, including VOT and vowel duration).
To ensure a fair comparison, for all models we estimated total articulation time using a manually-annotated window encompassing the full set of syllables.

Table \ref{tab:ddkcorrelations} presents the correlations and mean absolute errors between the model and the annotator. DDK rates predicted by our models were highly correlated with those of the annotator. Of the four models, the \lstm~model performed the best, achieving correlations of 0.94 and 0.99. It successfully predicts DDK rates from completely unannotated DDK samples in a way that generalizes across datasets and populations. 

\begin{table}[h]
\caption{\label{tab:ddkcorrelations}Correlations between model and annotator DDK rates (mean absolute errors in parentheses). The result on Younger NT Adults refers to the performance on the test set. Bold indicates the best performing model within each column. All correlations are significant with p $<$ 1e-5.} 
\resizebox{0.47\textwidth}{!}{%
\begin{tabular}{lccccccc}
\hline\hline
 & Younger NT Adults & Older PD Adults \\
\hline
\lstm~ & \textbf{0.94 (0.13)} & \textbf{0.99 (0.05)} \\
\cnn~ &   0.92 (0.22) & 0.85 (0.32) \\
Arias-Vergara et al. \cite{gru} &  0.88 (0.24) & 0.94 (0.22) \\
R\"{a}s\"{a}nen et al. \cite{rasanen2018pre} & 0.56 (0.86) & 0.93 (0.51) \\
Annotators & 1.00 (0.02) & 1.00 (0.00) \\
\hline\hline
\end{tabular}}
\vspace*{-3mm}
\end{table}

\subsubsection{Segment duration and boundaries}

Recall that we group frames with the same predicted object type to a single segment. The model performance are measured at the segment level, and we analyze how the accuracy of the predicted boundaries and the durations of both VOT and vowel segments. 

For computing performance of the model we created an assignment of prediction-target segments pairs as follows. For each of the predicted segments, we found the target segment with the closest start-time and end-time. We used it as the target assignment with the highest overlap. Miss-detection is considered for these target segments which are excluded in the overlapped region. False alarms were computed correspondingly.

We used the above assignment to compute F1. Table~\ref{tab:ddk_f1} presents F1-scores for VOT and vowel segments detection for each models by dataset (cf.~\cite{kreuk2020self}). It seems that the F1-scores for all models are high, while \lstm~shows the highest F1-score. These results suggest that there are not many miss detection and false positives.

We now turn to evaluate the accuracy on the boundaries and the durations. To do so, we need to remove all the miss detections and false positives of all models, and limit our analyses to segments that were identified by both the model and the annotator. Furthermore, we removed outliers (the top 5\% and bottom 2\% of duration values, chosen based on analyses on the validation set) from each of the models and annotator's predictions as suggested in \cite{goldrick2016automatic}.


\begin{table}
\caption{\label{tab:ddk_f1}F1-scores for VOT and vowel segments prediction.}
\resizebox{0.47\textwidth}{!}{%
\begin{tabular}{lccccccc}
\hline\hline
 & \multicolumn{2}{c}{Younger NT Adults} & \multicolumn{2}{c}{Older PD Adults}\\[0.05cm]
& VOT F1 & Vowel F1 & VOT F1 & Vowel F1 \\
\hline
\lstm & \textbf{0.978} &\textbf{0.985} &\textbf{0.993}&\textbf{0.998} \\
\cnn &   0.959 & 0.957& 0.974 &0.964 \\
Arias-Vergara et al. \cite{gru} &  0.945 & - & 0.950 & -\\
\hline\hline
\end{tabular}}
\end{table}

\begin{table}
\caption{\label{tab:segdurcorrelations}Correlations between model and annotator durations by dataset (mean absolute errors in parentheses). Bolded values represent the best-performing models within each column (VOT or Vowel in each dataset). All correlations are significant with p $<$ 1e-33.} 
\resizebox{0.47\textwidth}{!}{%
\begin{tabular}{lccccccc}
\hline\hline
 & \multicolumn{2}{c}{Younger NT Adults } & \multicolumn{2}{c}{Older PD Adults }\\ [0.05cm]
& VOTs & Vowels & VOTs & Vowels \\
\hline
\lstm~ & \textbf{0.90 (.004)} & \textbf{0.85 (.007)} & \textbf{0.65 (.006)} & \textbf{0.87 (.008)} \\
\cnn~ & 0.85 (.004) & 0.84 (.009) & 0.65 (.007) & 0.83 (.011) \\
Arias-Vergara et al. \cite{gru} & 0.80 (.006) & - & 0.50 (.008) & - \\
Annotators & 0.93 (.003) & 0.94 (.004) & 0.78 (.004) & 0.93 (.005) \\
\hline\hline
\end{tabular}
}
\vspace*{-3mm}
\end{table}

\begin{table}[t]
\caption{\label{tab:mad_boundaries}Mean absolute deviation in boundary offsets (milliseconds) by dataset.}

\resizebox{0.49\textwidth}{!}{
\begin{tabular}{llcccccccc}
\hline\hline

& Model & VOT  & VOT Offset/ & Vowel \\
&  & Onset & Vowel Onset & Offset\\
\hline
\multirow{4}*{\minitab[l]{Younger NT\!\!\!\!\!\!\!\!\\Adults}  } &\lstm & \textbf{1.88} & \textbf{2.96} & \textbf{6.34} \\
&\cnn~  & 1.97 & 3.17 & 7.38  \\
&Arias-Vergara et al. \cite{gru} & 3.86 & 4.51 & - \\
&Annotators & 1.10 & 2.30 & 2.62 \\
\hline
\multirow{4}{*}{\minitab[l]{Older PD\!\!\!\!\!\!\!\!\\ Adults} } &\lstm &  3.08 & 6.24 & \textbf{3.43}\\
&\cnn~ & \textbf{3.01} & 6.16 & 5.46 \\
&Arias-Vergara et al. \cite{gru} &  5.63 & \textbf{5.60} & - \\
&Annotators &  1.95 & 2.99 & 2.04 \\
\hline\hline
\end{tabular}
}
\vspace*{-3mm}
\end{table}

Table \ref{tab:segdurcorrelations} shows the correlation between the predicted duration and the annotated duration for each of the models, as well as the mean error rates. \lstm~achieves the highest correlations with the annotator (and lowest errors) across test sets, followed by \cnn~and then the model of Arias-Vergara et al. \cite{gru}, which shows the lowest correlations (and highest errors).

Table \ref{tab:mad_boundaries} presents the mean absolute deviation of boundary location across boundary types (VOT onset, VOT offset/vowel onset, and vowel offset) of the models and annotators (as reported in \cite{keshet2001plosive}). For the Younger NT Adults test set, the \lstm~model shows the best performance. It performs comparably to annotators for the VOT boundaries (deviations between 1.5-3 ms), but shows somewhat higher mean absolute deviations for the vowel offset ($\sim$6 ms). When evaluating on Older PD Adults dataset, the results are slightly worse and no single model outperforms the others across all boundary types. However, \lstm~ still shows good performance with mean absolute deviations ranging from around 3-6 ms. Overall, \lstm~ is able to reliably identify the boundaries of segments (and determine their durations), which will allow researchers and clinicians to automatically measure a host of segmental properties beyond DDK rate.


\section{Conclusions}
\label{sec:conclusions}

Automated analysis of fine-grained acoustic properties of diadochokinetic speech could provide new insights into speech motor disorders without increasing burdens on clinicians and researchers. In this paper, we presented two new algorithms for segmenting VOTs and vowels from unannotated DDK samples and compared their performance against the current state-of-the-art deep learning model \cite{gru} and signal proceesing model \cite{rasanen2018pre}. We evaluated the models in their ability to (i) predict DDK rate and (ii) identify duration and boundary location of consonats and vowels. We found that, in general,  \lstm~achieved better performance than \cnn, which suggests that recurrent neural network (RNN) cannot be entirely replaced by CNN for sequential tasks. Overall,  \lstm~ achieved state-of-the-art performance on unannotated speech, performing almost as well as human annotators across two datasets. 

These systems could allow for more nuanced, detailed, and objective measures of DDK samples. The temporal boundaries extracted here can inform many other acoustic analyses. These include: spectral analysis of VOTs and vowels (e.g., burst spectra, formant properties); and more detailed analysis of the temporal/metrical properties of production (e.g., variability in speech rate over a trial). As measurement is automatic, such analyses can be conducted over many trials and individuals, allowing for more detailed assessment of the distributional properties of measures.

There are several aspects of the model that can be improved in future work. As noted above, the model misses syllables, especially when the participant flaps /t/s. We have also noted poor performance on a small number of individuals with high degrees of creaky phonation. More extensive training on these less frequent acoustic variants (flaps, creaky voice) may improve performance. In addition, this model was applied only to voiceless targets; future work can extend this approach to examine AMR and SMR subtasks using voiced variants. Finally, the model's robustness to variation in recording conditions should be examined (e.g., by analysis of speech collected outside of acoustically controlled laboratory environments).

In conclusion, we have introduced a deep neural network model, \lstm, which reliably extracts clinically useful information from completely untranscribed and unannotated DDK samples. This algorithm can allow for more detailed automatic analyses of DDK samples, providing new insights into motor speech behavior.

\section{Acknowledgment}
\label{sec:acknowledgment}

This work is supported by the Ministry of Science \& Technology, Israel (Y. Segal); U.S. National Institutes of Health (NIH; grants R21MH119677, K01DC014298, R01DC018589); and the Ontario Brain Institute with matching funds provided by participating hospitals, the Windsor/Essex County ALS Association and the Temerty Family Foundation. The opinions, results, and conclusions are those of the authors and no endorsement by the Ontario Brain Institute or NIH is intended or should be inferred. Thanks to Hung-Shao Cheng, Rosemary Dong, Katerina Alexopoulos, Camila Hirani, and Jasmine Tran for help in data collection and processing.

\bibliographystyle{IEEEtran}

\bibliography{mybib}

\end{document}